\documentclass{article}

\usepackage{graphicx}
\usepackage[preprint]{aaj2026}
\usepackage{natbib}

\usepackage[utf8]{inputenc} 
\usepackage[T1]{fontenc}    
\usepackage{hyperref}       
\usepackage{url}            
\usepackage{booktabs}       
\usepackage{amsfonts}       
\usepackage{nicefrac}       
\usepackage{microtype}      
\usepackage{xcolor}         
\usepackage{amsmath}
\usepackage{ascmac}

\title{Extended predictive coding framework as variational free-energy minimisation under exponential-family assumption}

\author{%
  Asaki Kataoka \\
  Neural Computation Unit\\
  Okinawa Institute of Science and Technology\\
  Okinawa, Japan \\
  \texttt{asaki.kataoka@oist.jp} \\
\And
 Kenji Doya \\
 Neural Computation Unit \\
 Okinawa Institute of Science and Technology \\
 Okinawa, Japan \\
 \texttt{doya@oist.jp}
}

\begin{document}

\maketitle

\begin{abstract}
    The sensory cortices of the brain perform perceptual inference efficiently through their complex networks of neurons. One of the theoretical accounts of this process is the free-energy principle (FEP), which postulates that the brain performs variational Bayesian inference. Pioneering studies have shown that FEP can correspond to the predictive coding (PC) hypothesis under the Gaussian assumption and Laplace approximation. However, PC-based implementations of FEP within such a limited Gaussian regime have failed to capture several properties of biological neural networks, such as nonlinearity and heterogeneity of input--output properties within a network, and the biological implausibility of negative firing rates. This study shows that, when a broader class of probability distributions, namely the exponential family of distributions (EFD), is assumed for the variational posterior and prior, these missing characteristics are exhibited within the network, maintaining the FEP--PC correspondence up to the second cumulant of the posterior. We also show that the proposed model can be trained by biologically plausible local plasticity rules. Our results enrich the explanatory power of FEP regarding neural dynamics involved in perception as variational inference.
\end{abstract}

\section{Introduction}
The free-energy principle (FEP), which has attracted immense attention from the neuroscience community, posits that the brain functions as a variational Bayesian estimator \cite{Friston2009-xb,Friston2009-vn,Friston2010-ya,Bogacz2017-al}. This theory is grounded in Helmholtz's early proposal of unconscious inference; since perception is fundamentally an ill-posed problem of inferring the hidden state of the external world from noisy, information-discarding sensory observations, the brain must implicitly infer it based on past experiences \cite{Helmholtz1867-fr}. This concept was later mathematically formalised as the Bayesian brain hypothesis, where prior experiences correspond to a prior distribution used to estimate the posterior \cite{Knill1996-jk, Doya2007-yy, Bottemanne2025-ji}. Within Marr's three-level notion, FEP serves as a principle at the computational level. It circumvents the intractable integration required for the marginal likelihood (or evidence) by stating that the brain instead minimises variational free energy---an upper bound on the negative log marginal likelihood, \textit{i.e.} the sensory surprisal.

Following the proposal of this theory by Friston, it has been extensively demonstrated that, under Gaussian assumption and/or Laplace approximation, a possible neural implementation of FEP takes a form structurally identical to predictive coding (PC) \cite{Friston2009-vn, Bogacz2017-al}. This was a foundational result, as it provided a plausible algorithmic and neural basis for this abstract principle. PC postulates two distinct types of neurons: \textit{representational neurons}, which encode current belief about the state of the external world and generate top-down predictions regarding lower-level activities, and \textit{error-coding neurons}, which compare these top-down predictions with bottom-up signals to send prediction errors back up the functional hierarchy \cite{Rao1999-uv}.

However, it has been unknown whether this correspondence holds for non-Gaussian assumptions. Previous studies on FEP relating it to PC have assumed that posterior, prior, and likelihood are all Gaussian, and the current literature has only shown this relation under such a limited assumption. This has created a huge gap between the general and broad explanatory power of FEP expected from its fundamental notion and the narrowness of assumption it requires for formulating promising neural implementation.

This study shows that the Gaussian assumption for the prior and posterior can be relaxed without drastically violating the FEP--PC correspondence. In particular, we show that the gradient descent dynamics can take the form of PC even when assuming that the posterior and prior belong to exponential family of distributions (EFD), which is a broader class of probability distributions beyond Gaussian regime. Under the EFD assumption for posterior and prior and Gaussian assumption for likelihood, the inferential dynamics of the representational neurons are driven by two additive signals: attraction drive towards the prior and prediction-error drive.

Based on this theoretical result, we propose \textit{EFD--FEP} model as an extended PC model. In each layer of this model, the approximate posterior is represented by a population of representational neurons. We define ``internal states'' of neurons, which correspond to the parameters of the posterior, and sampled random values corresponding to the firing activities of these neurons. The prior-attracting drives and prediction signals are represented by top-down message-passing, and the additive prediction errors are propagated in a bottom-up manner.

Furthermore, our model reflects several known biological and electrophysiological properties. These features enhance the explanatory power of FEP as computational formulation of neural computation involved in perceptual inference in the brain.

First, synaptic weight updates, also derived from gradient descents in the shared objective function as the inferential dynamics, further enhance the plausibility of its biological neural implementation. The fundamental learning rules of our model are twofold (Section~\ref{sec:learning}): Hebbian-like plasticity and a more complex plasticity rule based on total input intensity. Although the latter might appear difficult to implement in biological neural systems at first glance, it conceptually aligns well with recent findings of plasticity rules based on dendritic plateau potentials and calcium-mediated firing (BAC firing) in the apical tufts of pyramidal neurons \cite{Larkum1999-no,Larkum2013-tz,Bittner2015-mp}.

Second, our model, particularly under the condition that the posterior can be factorised into independent elementwise distributions, allows for heterogeneity or diversity and nonlinearity in neuron-level electrophysiological frequency--current (F--I) properties, or activation functions, within a single network, having been observed experimentally \cite{Shamir2006-uh,Padmanabhan2010-el, Angelo2011-jy, Ciarleglio2015-et, Moradi-Chameh2021-wy}. Previous PC models, especially those normatively derived from Gaussian modelling \cite{Rao1999-uv, Friston2009-vn,Bogacz2017-al,oliviers_learning_2024}, have assumed linear activations; these models require neurons to ``fire'' at firing rates proportional to the intensity of the input. Unlike such models, our framework accepts arbitrary F--I curves as long as they satisfy only two conditions: monotonically increasing and differentiable. This also resolves the negative-firing problem that has often been observed in previous PC models. While linear activations allow the model to sample negative firing activities, biological neurons do not emit negative firing rates. Although introducing rectification can suppress such negativity, this modification is non-normative and fundamentally violates VFE minimisation in that regime. Conversely, our model accepts nonlinear activations that converge to, but do not fall below, zero, thereby preventing negative firing while maintaining the normative objective of VFE reduction. Furthermore, our framework embraces the heterogeneity of nonlinearity. This allows for, for example, neuron-wise difference of shifts and gains of the sigmoidal activation, or coexistence of neurons with sigmoidal, exponential, and hyperbolic activations within a single population of neurons.

Unlike the previous PC models and Gaussian-assuming FEP framework on which they have relied, our model keeps the PC framework while ensuring that it effectively reduces the VFE as postulated in FEP. In particular, the biology-aware properties found in our model further strengthen FEP and PC hypothesis as an explanatory theory for neural computations in the brain and its cortices. We also discuss several remaining issues regarding biological plausibility left even in our model.

\section{Background}
\subsection{free-energy principle}
The brain infers the state of the world using limited information provided by sensory inputs. For example, inferring the spatial arrangement of two objects from two-dimensional retinal projections is theoretically unsolvable. Even when one object is in front of another and occluding it, it cannot be determined with certainty whether one object is truly in front, or if they are at the same distance but one object has a missing part that perfectly mimics the shape of occlusion. However, the brain rarely falls into such unnatural estimations, and we confidently perceive the former, more plausible situation.

A key idea for understanding how the brain solves such ill-posed problems is \textit{unconscious inference}, first proposed by von Helmholtz in the 19th century \cite{Helmholtz1867-fr}. This concept states that the brain must infer the state of the world based on prior beliefs about what the world is likely to look like, which are constructed or learnt from past experiences. The result of this inference explains the sensory input in a way that is consistent with these prior beliefs, for example, favoring the interpretation that ``one object is in front of another'' over ``one object has a missing part'', unless learning fails and results in an unnatural prior belief.

Later, this idea was mathematically formalised as the Bayesian brain hypothesis \cite{Knill1996-jk, Doya2007-yy, Bottemanne2025-ji}. This hypothesis posits that the brain learns a \textit{generative model} $p(x, s)$ of the sensory input $s$ and the latent variable $x$, which represents states of the world inaccessible to direct observation. Given the generative model, the probability distribution of $x$ conditioned on the sensory input, namely the \textit{posterior distribution}, can be computed via Bayes' rule:
\begin{equation*}
    p(x|s)=\frac{p(x, s)}{p(s)}.
\end{equation*}

A major difficulty of the Bayesian brain hypothesis is that the exact neural substrates for storing the generative model and implementing Bayesian computation remain ambiguous. Storing the co-occurrence probabilities of all possible latent states and sensory inputs is impossible, as the full repertoire of $x$ is unknown to the brain. Additionally, it is often argued to be implausible that the brain computes the marginal likelihood $p(s)=\int p(x, s)dx$. This would require a rapid integration of the likelihood over all possible values of $x$. These challenges have led to the exploration of alternative neural computations that approximate Bayes' rule.

The free-energy principle (FEP), proposed by Friston, provides a more implementable framework for the Bayesian brain by introducing variational Bayesian inference \cite{Friston2009-xb, Friston2010-ya, Bogacz2017-al}. According to FEP, instead of directly calculating the posterior using Bayes' theorem, the brain seeks to find the parameter $\vartheta$ of the \textit{approximate posterior distribution} $q$ that minimises the \textit{variational free energy} (VFE), defined as:
\begin{align}
    \mathcal F &= \mathbb E_q\left[\log\frac{q(x; \vartheta)}{p(x|s)} - \log p(s)\right] \label{eq:vfe_as_post_distance} \\
    &= \underbrace{D_\text{KL}[q(x; \vartheta)\,\|\,p(x; m)]}_{\text{KLD term}} - \underbrace{\mathbb E_q[\log p(s|x; \phi)]}_{\text{Log-likelihood term}} \label{eq:vfe_normal_form}.
\end{align}
The first formulation (Eq.~\ref{eq:vfe_as_post_distance}) reflects a fundamental concept of this objective function; for a fixed sensory input $s$, the VFE corresponds to the Kullback--Leibler divergence (KLD) between the approximate posterior and the \textit{true} posterior, plus a constant (the negative log evidence). This form explicitly shows that $\vartheta$ should be adjusted to minimise this KLD. However, a remaining challenge is that the brain does not have access to the true posterior $p(x|s)$. Consequently, the prior $p(x; m)$ and the likelihood $p(s|x; \phi)$ are modelled as being parameterised by $m$ and $\phi$, respectively. By parameterising all three distributions, the posterior $q$, the prior $p_m$, and the likelihood $p_\phi$, the brain no longer needs to store complete probability distributions. Instead, by assuming a specific family of distributions (\textit{e.g.} Gaussian), these finite-dimensional parameters can represent the distributions. By fixing $s$ or averaging $\mathcal F$ over all possible values of $s$, the VFE becomes a function of $(\vartheta, m, \phi)$. FEP typically explains perception and learning as the minimisation of the (average) VFE with respect to $\vartheta$ and $(m, \phi)$, respectively\footnote{Another attempt of Friston and colleagues is to include action planning and decision-making of actions into account of FEP. A major theoretical advances have been made as \textit{active inference}\cite{Friston2014-za,Kirchhoff2018-mq,Parr2021-xd}. However, it is noteworthy that (or at least we understand that) the theoretical formulations of them need to take ``expected free energy'' into account and is not strictly identical to the VFE formulation (Eq.~\ref{eq:vfe_normal_form}). In this study, we do not argue active-inference formulations and consider the aspect of FEP as a variational formulation of perceptual inference.}.

Note that the adjustment of these parameters, especially $\vartheta$, is not performed in a one-shot manner (\textit{i.e.} amortised inference), as is common in many artificial neural network models. Instead, within FEP framework, these parameters are gradually updated towards their optimal values as they descend the landscape of the VFE toward a minimum.

\subsection{Gaussian Assumption and Predictive Coding}
One of the major neural implementations of FEP is derived from the Gaussian assumption, which leads to neural message-passing in the form of predictive coding\footnote{There has been, to be more precise, other attempts to explore possible neural implementations and related assumptions proposed by them. For example, previous studies have attempted to introduce the idea of Markov blanket and generalised coordinate systems \cite{Friston2009-vn,Friston2013-ii,Kirchhoff2018-mq}. While these additional assumptions lead us to interesting implications about possibility of neural implementations of FEP, we take a simpler stance on interpretation of what FEP itself covers and do not explicitly take such assumptions into our considerations; the major statement of FEP is simply that the brain achieves perceptual inference by reducing the VFE. In fact, even under these assumptions, it is general that their mathematical formulations of implementation of FEP typically rely on the Gaussian assumptions to some degrees.} \cite{Friston2009-vn, Bogacz2017-al}. Predictive coding (PC) is a hypothesis of neural computation which postulates that there are two types of functionally distinct neurons in the cortex: \textit{representational} neurons and \textit{error-coding} neurons \cite{Rao1999-uv}. In PC theory, the representational neurons are considered to encode the current estimate of the state of the world. Predictions regarding the activity of representational neurons in hierarchically lower layers are generated from these neurons via synaptic connections. The error neurons in the layer below compare the actual lower-layer representational activity with the prediction and send the resulting error back up the hierarchy. The representational neurons then update and modify the current inference based on this error signal. While Friston and his colleagues have extensively explored the neural implementations of VFE minimisation under various assumptions, the most standard implementation emerges from the Gaussian assumption. A key characteristic of this derivation is that the result implies neural processing that corresponds directly to PC.

Here we briefly summarise the derivation of PC-based neural processing from the Gaussian FEP. Let the prior and the likelihood be Gaussian distributions \cite{Bogacz2017-al}:
\begin{align*}
    p(x; m) &= \mathcal{N}(x; \mu_m, \Sigma_m), \\
    p(s|x; \phi) &= \mathcal{N}(s; \mu_\phi(x), \Sigma_\phi).
\end{align*}
For the approximate posterior, a common assumption introduced by Friston and colleagues is the Laplace approximation, where the distribution is sharply concentrated around its mean:
\begin{equation*}
    q(x; \vartheta) = \delta(x - \mu_\vartheta).
\end{equation*}
Under these assumptions, the expectation calculation simplifies, and the VFE is evaluated as:
\begin{align*}
    \mathcal{F} &= -\log p(\mu_\vartheta; m) - \log p(s|\mu_\vartheta; \phi) \notag \\
    &= \frac{1}{2} \left( \log |\Sigma_m| + \frac{\|\mu_m - \mu_\vartheta\|_2^2}{\Sigma_m} + \log |\Sigma_\phi| + \frac{\|s - \mu_\phi(\mu_\vartheta)\|_2^2}{\Sigma_\phi} \right) + \text{const.}
\end{align*}
By calculating the derivative of the VFE with respect to $\mu_\vartheta$, we obtain the temporal update rule for $\mu_\vartheta$ given a sensory input:
\begin{equation}
    \dot{\mu}_\vartheta \propto \frac{\partial \mathcal{F}}{\partial \mu_\vartheta} = \frac{\mu_m - \mu_\vartheta}{\Sigma_m} + \frac{s - \mu_\phi(\mu_\vartheta)}{\Sigma_\phi} \mu_\phi'(\mu_\vartheta) \label{eq:gaussian_fep_result}.
\end{equation}
The first term represents the drive to maximise the agreement between the prior belief ($\mu_m$) and the current estimate ($\mu_\vartheta$). The second term can be interpreted as weighted error feedback, where $\mu_\phi(\mu_\vartheta)$ acts as a prediction of the sensory input generated from the current estimate of the latent state. This implies neural dynamics where $\mu_\vartheta$ corresponds to the activity of representational neurons and $s - \mu_\phi(\mu_\vartheta)$ corresponds to the activity of error-coding neurons, particularly when $\Sigma_m$ and $\Sigma_\phi$ are diagonal. This is the simplest form of the correspondence between FEP and PC theory. One of the theoretical advantages of FEP is that it can be implemented in a manner consistent with the widely recognised PC hypothesis.

\subsection{When FEP--PC Correspondence Holds}
It is a common misconception that FEP is identical to PC theory. FEP only postulates that the brain infers the state of the world by reducing the VFE, whereas PC theory defines neural processing based on the propagation of prediction-error feedback signals. While various neural implementations of FEP have been proposed, most have relied on the Gaussian assumption and have only demonstrated that FEP-PC correspondence holds within that specific regime.

In fact, under certain assumptions, FEP does not lead to PC-like neural processing. An extreme example is the Cauchy--Lorentz distribution, where the KLD term cannot be analytically evaluated in closed form. This would require the brain to compute the integral over all possible values of $x$ (or at least an approximation thereof), which is typically considered a computationally intractable task for biological systems.

Given such a variability of the characteristics of derived inferential dynamics, a question would be when free-energy minimisation are written in the form of predictive coding. Although the log-likelihood term in the VFE (Eq.~\ref{eq:vfe_normal_form}) intuitively corresponds to ``error'', it is not trivial that the gradient of this term always can be expressed as additive error feedback as in previous Gaussian models (Eq.~\ref{eq:gaussian_fep_result}). To understand whether and how neural systems solve the Bayesian problem, clarifying the relationship between assumptions or constraints introduced to a principle and the properties of the resulting inferential frameworks is significant.

A partial answer provided by this study is that the Gaussian assumption for the posterior and prior can be relaxed and we can assume a broader class of probability distributions, namely the EFD. This relaxation does not break the correspondence between PC-based dynamics and VFE reduction, up to the second-order posterior cumulant, but introduces its neural implementations reflecting several known biological properties observed in the neural systems. While the model we present in this article does not relax the Gaussian assumption on the likelihood function, we also discuss the case of non-Gaussian EFD likelihoods to argue the condition where the PC-based inferential dynamics no longer performs as an effective variational inference (Appendix~\ref{apsec:nongaussian_likelihood}).

\section{Results: Theoretical derivation of the EFD--FEP model\label{sec:results}}
\begin{figure}
    \centering
    \includegraphics[width=\linewidth]{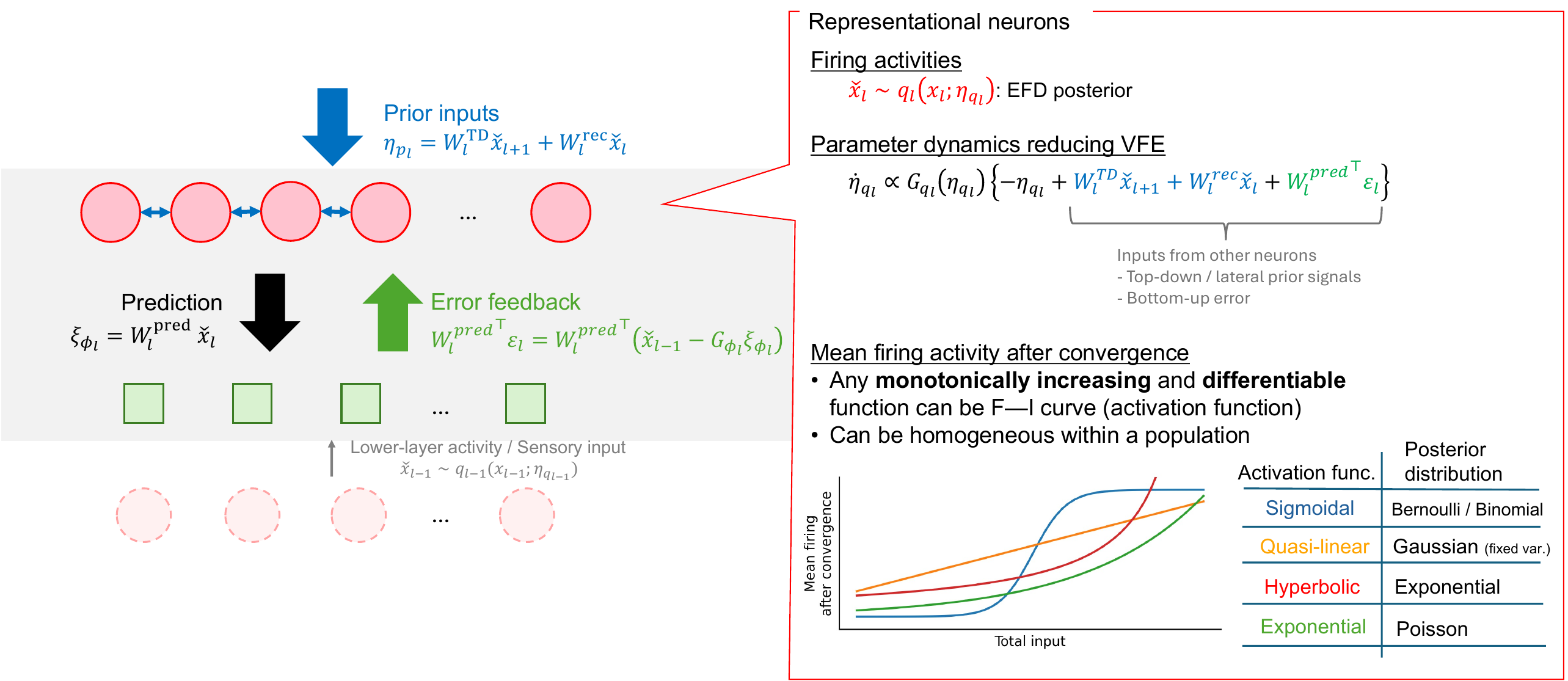}
    \caption{Schematic illustration of EFD--FEP model derived in Section~\ref{sec:results}. Representational neurons (left, red balls) with internal states $\eta_{q_l}$ emit firing activities (e.g., binary spike, spike count within time window, or firing rate) $\check x_l$ sampled from the EFD posterior $q_l$ parameterised by $\eta_{q_l}$. Sampled activities are transformed into prediction about the lower-layer activitiese (left, black arrow). $\eta_{q_l}$'s temporal evolution along approximate VFE gradient---derived in Sections~\ref{sec:derivation} and \ref{sec:stochastic}---is driven by the prediction error fed back up the hierarchy (left, green arrow) and prior signals propagated via lateral and another top-down pathway (left, blue arrows). Neuron-wise mean firing activity $\mathbb E_{q_l}[\check x_l]$ after convergence given constant input is characterised by activation function $\nabla A_q(\text{Total input})$. Any monotonically increasing and differentiable functions can be the activation function. The left-bottom panel shows examples of correspondence between the activation function and class of the posterior; sigmoidal--Bernoulli/binomial, linear--Gaussian, hyperbolic--exponential, and exponential--Poisson (see Section~\ref{sec:actfunc}).}
    \label{fig:model_schematic}
\end{figure}

In this section, we theoretically derive our novel \textit{EFD--FEP model}, a generalised predictive coding model based on FEP under the assumption of the exponential family. The schematics of the derived model is shown in Fig.~\ref{fig:model_schematic}. After clarifying the prerequisites for the derivation, we show that gradient descent on VFE with EFD-assumed prior and posterior and Gaussian-assumed likelihood generally takes the form of predictive coding as we neglect the third posterior cumulant. We also introduce learnable parameters regulating the VFE landscape and show that their learning rules, also derived as VFE descent, can be expressed in a simple form. 

After deriving the general FEP--PC correspondence across the EFD regime in Section~\ref{sec:derivation}, we discuss the mapping of these mathematical variables onto biological substrates. In particular, within a specific subtype of the model (see Section~\ref{sec:subtyping}), each component of the posterior parameters $\eta$ and $\mu$ maps onto the internal states of a neuron. Specifically, these correspond to dynamical properties akin to membrane potentials and firing rates, respectively, while $\check{x}$ may represent generated spike signals or their counts. Furthermore, we argue that the characteristics of the log-partition function $A$ regulate the electrophysiological properties of a neuron, such that its gradient $\nabla A$ correlates with the frequency--current (F--I) curve.

\subsection{Prerequisites}
\subsubsection{Hierarchical factorisation\label{sec:factorisation}}
We assume a \textit{hierarchical} form of FEP. Let $x$ be the latent variables of the world, which are inaccessible to the brain. We assume that $x$ is an ensemble of multiple vectors, \textit{i.e.} $x=[x_1, \dots, x_L]$, and that the generative model $p(x, s)$ can be factorised into the following Markov chain:
\begin{equation*}
    p(s, x) = p(s|x_1) \prod_{l=1}^{L-1} p(x_{l}|x_{l+1}) p(x_L).
\end{equation*}
Given this factorisation, we also define the approximate posterior as a product of sub-posteriors defined independently for each layer:
\begin{equation*}
    q(x; \vartheta) = \prod_{l=1}^L q(x_l; \vartheta_l).
\end{equation*}
Under this assumption, we define the total VFE as the sum of layer-wise VFEs as follows:
\begin{equation}
    \mathcal{F} = \sum_l \mathcal{F}_l = \sum_l \left\{ D_\text{KL}[q(x_l; \vartheta_l) \,\|\, p(x_l; m_l)] - \mathbb{E}_{q(x_l; \vartheta_l)}[\log p(x_{l-1}|x_l; \phi_l)] \right\}. \label{eq:factorised}
\end{equation}
This is equivalent to stating that the representational neurons in each layer infer the state of those in the layer below by observing their activities (\textit{e.g.} emitted spikes).

\subsubsection{The exponential family of distributions\label{sec:efd}}
The \textit{exponential family of distributions} (EFD) is a class of probability distributions whose probability density or mass functions take the following form:
\begin{equation}
    f(x; \eta) = h(x) \exp[\eta^\top T(x) - A(\eta)].
\end{equation}
Here, $\eta$ represents the \textit{natural parameters} of the distribution, and $T(x)$ denotes the \textit{sufficient statistics}. Intuitively, $T(x)$ corresponds to the minimal statistics whose sum is sufficient to determine the full shape of the distribution. For instance, in cases such as the Bernoulli and Poisson distributions, where the controlling parameters are aggregated into a single value (typically the mean parameter), the sufficient statistic is $T(x) = x$. Conversely, for distributions such as the Gaussian, where second-order statistics are also required for determination, the sufficient statistics are $T(x) = [x, xx^\top]^\top$. For ease of notation, we hereafter denote the sufficient statistics as $\check{x} := T(x)$. The term $A(\eta) := \log \int h(x) \exp[\eta^\top T(x)] dx$ is referred to as the \textit{log-partition function}.

We note several important characteristics of the EFD. First, the log-partition function $A(\eta)$ is strictly convex and twice differentiable. Formally, this is equivalent to stating that the real parts of all eigenvalues of the Hessian $\nabla^2 A(\eta)$ are positive for any $\eta$ (i.e. the Hessian is positive definite). It is also important to note that this Hessian
\begin{equation*}
    G(\eta):=\frac{\partial \mu}{\partial\eta}=\nabla_\eta^2
    A(\eta)
\end{equation*}
is Fisher information matrix, and particularly since $p$ belongs to EFD, also equal to covariance matrix. 

We refer to the first-order gradient of the log-partition function,
\begin{equation}
    \mu := \nabla A(\eta), \label{eq:mu_from_eta}
\end{equation}
as the \textit{Legendre-dual} parameter or the \textit{mean} parameter. Indeed, it is a known property of the EFD that $\mu = \mathbb{E}[\check{x}]$. Crucially, the convexity of $A$ guarantees that the mean parameter is a monotonic mapping of $\eta$. Due to this property, there exists an inverse mapping from $\mu$ back to $\eta$, given by
\begin{equation}
    \eta = \nabla A^*(\mu), \label{eq:eta_from_mu}
\end{equation}
where $A^*(\mu) := \mu^\top \eta - A(\eta)$ is the \textit{Legendre transform} of $A(\eta)$, which is equivalent to negative entropy of the distribution.

Using the Legendre-dual parameters and the Legendre transformation, the KLD between two distributions belonging to the EFD simplifies to a remarkably concise form. Consider two distributions with the following probability densities:
\begin{align*}
    f(x; \eta_A) &= h(x) \exp[\eta_A^\top \check{x} - A(\eta_A)], \\
    g(x; \eta_B) &= h(x) \exp[\eta_B^\top \check{x} - B(\eta_B)].
\end{align*}
The KLD between these two distributions is given by:
\begin{equation}
    D_\text{KL}[f\,\|\,g] = A^*(\mu_A) + B(\eta_B) - \eta_B^\top \mu_A. \label{eq:KLD_EFD}
\end{equation}
This expression is particularly useful when evaluating the VFE, as the first term of the VFE contains the KLD between the approximate posterior and the prior distributions.

\subsection{Extended predictive coding as a form of VFE reduction for EFD posterior and prior}
In this section, we show the core theoretical result of this study. Unlike previous studies, we put the EFD assumptions for posterior and prior distributions, which is broader class of probability distributions beyond Gaussian which has often been assumed in previous studies. Section~\ref{sec:derivation} shows the theoretical derivation of the inferential dynamics, followed by Section~\ref{sec:stochastic} introducing its stochastic variant. Then, Section~\ref{sec:learning} shows that the learning mechanisms of the synaptic parameters, which is derived as a gradient descent of the shared objective function, namely VFE, w.r.t. them can be expressed as biologically-plausible local plasticity rules.

\subsubsection{Inferential dynamics as deterministic gradient descent\label{sec:derivation}}
Under the hierarchical factorisation introduced in Section~\ref{sec:factorisation}, the VFE is decomposed into layer-wise components, as shown in Eq.~\ref{eq:factorised}. We now assume that approximate posterior, prior, and likelihood all belong to the EFD:
\begin{align*}
    q_l(x_l; \eta_{q_l}) &:= h_{x_l}(x_l) \exp[\eta_{q_l}^\top \check{x}_l - A_{q_l}(\eta_{q_l})], \\
    p_l(x_l; \eta_{p_l}) &:= h_{x_l}(x_l) \exp[\eta_{p_l}^\top \check{x}_l - A_{p_l}(\eta_{p_l})], \\
    p_{\phi_l}(x_{l-1}|x_l; \phi_l) &:= h_{x_{l-1}}(x_{l-1}) \exp[\xi_{\phi_l}(\check{x}_l)^\top \check{x_{l-1}} - A_{\phi_l}(\xi_{\phi_l})].
\end{align*}
Note that the natural parameter $\xi_{\phi_l}$ of the likelihood function is determined as a function of the sufficient statistics $\check x_l$ sampled from the posterior $q_l$. This makes the likelihood $p_{\phi_l}$ behave as a random variable whose distribution is coupled with the posterior. Here, we assume that the posterior and prior can take arbitrary forms as long as the base measure $h_{x_l}$ is shared, while the likelihood function is fixed-variance Gaussian, which is represented as
\begin{equation}
A_{\phi_l}=\frac{1}{2}\xi_{\phi_l}^\top G_{\phi_l}\xi_{\phi_l},\label{eq:quadratic_logpart}
\end{equation}
where $G_\phi$ corresponds to the covariance matrix.

For these distributions, we assume that $h_\bullet$ and $A_\bullet$ are fixed, and each distribution is adjustable only through changes in its natural parameters $\eta_\bullet$ or $\xi_{\phi_l}$. Note that, in the general case, the dimensionality $d_l$ of the natural parameter space ($\eta_{q_l}, \eta_{p_l}\in\mathbb R^{d_l}$) does not necessarily equal the dimensionality of the random variable $x_l$. Rather, the natural parameters share the same dimensionality as the sufficient statistics $\check{x}_l$.

Given these assumptions, the VFE of the $l$-th layer can be expressed in the following form:
\begin{align}
    \mathcal{F}_l &= A^*_{q_l}(\mu_{q_l}) + A_{p_l}(\eta_{p_l}) - \eta_{p_l}^\top \mu_{q_l} - \mathbb{E}_{q_l} [\xi_{\phi_l}(\check{x}_l)^\top \check{x}_{l-1}] + \mathbb{E}_{q_l} [A_{\phi_l}(\xi_{\phi_l}(\check{x}_l))] - \log h_{x_{l-1}} \notag\\
    &= \underbrace{A_{q_l}^*(\mu_{q_l})+A_{p_l}(\eta_{p_l})-\eta_{p_l}^\top\mu_{q_l}}_{\text{KLD terms}}\underbrace{-\bar\xi_{\phi_l}^\top\check x_{l-1}+\frac{1}{2}\bar\xi_{\phi_l}^\top G_{\phi_l}\bar\xi_{\phi_l}+\frac{1}{2}\text{Tr}[G_{\phi_l}\Sigma_{\xi_{\phi_l}}]-\log h_{x_{l-1}}}_{\text{Log-likelihood terms}}.\label{eq:EFD_VFE}
\end{align}
Here, we denote the covariance matrix of $\xi_{\phi_l}(\check x_l)$ as $\Sigma_{\xi_{\phi_l}}$, and $q_l$-mean of $\xi_{\phi_l}(\eta_{q_l})$ as $\bar\xi_{q_l}:=\mathbb E_{q_l}[\xi_{\phi_l}(\check x_l)]$. Regarding the KLD term, we have applied the general expression for EFDs given in Eq.~\ref{eq:KLD_EFD}. Note that the expectation $\mathbb{E}_{q_l}$ is taken with respect to the approximate posterior of the $l$-th layer.

The formulation above represents the general form of the VFE under the EFD assumption and hierarchical decomposition. For formalising the biological neural implementation of the derived model, we introduce an assumption of linear coupling from the posterior to the likelihood
\begin{equation*}
    \xi_{\phi_l}(\check{x}_l) := W_l^\text{pred} \check{x}_l,
\end{equation*}
where $W_l^\text{pred} \in \mathbb{R}^{d_{l-1} \times d_l}$ is a linear operator, which will later be mapped on synaptic weights. The sampled sufficient statistics $\check x_l$ is transformed into a prediction of the lower-layer states via this rule. By the linearity, this also introduces $\bar\xi_{\phi_l}=W_l^\text{pred}\mu_{q_l}$ and $\Sigma_{\xi_\phi}=W_l^\text{pred}G_{q_l}(\eta_{q_l})W_l^{\text{pred}~\top}$.

Now, we derive the dynamics of $\eta_{q_l}$ in a direction that minimises the VFE, which corresponds to the negative gradient of the VFE with respect to $\eta_{q_l}$. Note that we will introduce an approximation for the gradient of log-partition term $-\mathbb E_q[A_\phi]$ in which the third-order cumulant of the posterior is neglected. Consequently, our derived dynamics take the following form:
\begin{equation*}
    \tau_\text{inference} \dot{\eta}_{q_l} \approx -\nabla_{\eta_{q_l}} \mathcal{F}_l.
\end{equation*}

First, we evaluate the gradient of the KLD terms. Using the coordinate transformation introduced in Eq.~\ref{eq:mu_from_eta} and chain rule, we obtain
\begin{align}
    -\nabla_{\eta_{q_l}}D_\text{KL}[q_l\|p_l]&=-\nabla_{\eta_{q_l}}\left\{A_{q_l}^*(\mu_{q_l})+A_{p_l}(\eta_{p_l})-\mu_{q_l}^\top\eta_{p_l}\right\} \notag\\
    &=G_{q_l}(\eta_{q_l})\left\{-\nabla_{\mu_{q_l}}A_{q_l}^*(\mu_{q_l})+\nabla_{\mu_{q_l}}\mu_{q_l}^\top\eta_{p_l}\right\} \notag\\
    &=G_{q_l}(\eta_{q_l})\big\{-\eta_{q_l}+\eta_{p_l}\big\}\label{eq:kl_grad}.
\end{align}

The gradient of the log-likelihood terms $L$ can also be evaluated analytically as follows:
\begin{align}
    -\nabla_{\eta_{q_l}}L&=\nabla_{\eta_{q_l}}\bar\xi_{\phi_l}^\top\check x_{l-1}-\frac{1}{2}\nabla_{\eta_{q_l}}\bar\xi_{\phi_l}^\top G_{\phi_l}\bar\xi_{\phi_l}+\frac{1}{2}\nabla_{\eta_{q_l}}\text{Tr}[G_{\phi_l}\Sigma_{\xi_{\phi_l}}]-\log h_{x_{l-1}}\notag \\
    &=G_{q_l}(\eta_{q_l})W_l^{\text{pred}\,\top}\check x_{l-1}-G_{q_l}(\eta_{q_l})W_l^{\text{pred}\,\top} G_{\phi_l}\bar\xi_{\phi_l}+\frac{1}{2}\nabla_{\eta_{q_l}}\text{Tr}[G_{\phi_l}\Sigma_{\xi_{\phi_l}}].\label{eq:ll_grad}
\end{align}
By defining the additive error $\bar\varepsilon_l:=\check x_{l-1}-G_{\phi_l}\bar\xi_{\phi_l}$ and neglecting the third posterior cumulant $\nabla_{\eta_{q_l}}G_{q_l}(\eta_{q_l})=\nabla_{\eta_{q_l}}^3A_{q_l}(\eta_{q_l})$\footnote{We consider that this neglection is valid to some extent. Intuitively, at least, the third cumulant $\nabla^3A_q$ tends to be sufficiently small when the first cumulant $\nabla A_q$ has bounded supports. For example, when the posterior is Bernoulli or Binomial, where the first cumulant is sigmoidal, the third cumulant takes smaller values compared to the first and second cumulants.}, we obtain the following approximation of the log-likelihood gradient:
\begin{equation}
    -\nabla_{\eta_{q_l}}L\approx G_{q_l}(\eta_{q_l})W_l^{\text{pred}\,\top}\bar\varepsilon_l. \label{eq:ll_grad_approx}
\end{equation}  

By combining Eqs.~\ref{eq:kl_grad} and \ref{eq:ll_grad_approx}, we obtain the total approximate gradient of the VFE, representing the ideal inference dynamics for $\eta_{q_l}$:
\begin{itembox}[l]{Inferential dynamics (ordinary gradient descent)}
\begin{equation}
    \tau_\text{inference} \dot{\eta}_{q_l} = G_{q_l}(\eta_{q_l}) \left\{ -\eta_{q_l} + \eta_{p_l} + W_l^{\text{pred}~\top} \bar{\varepsilon}_l \right\} \approx -\nabla_{\eta_{q_l}} \mathcal{F}_l. \label{eq:ogd_inference}
\end{equation}
\end{itembox}
This form of the approximate gradient holds for any distribution belonging to the EFD, regardless of the specific functional form of its probability density or mass function, as long as we focus on the second and lower cumulant of the posterior. We refer to models following these dynamics as \textit{EFD--FEP models}.

An important feature of the inference dynamics in Eq.~\ref{eq:ogd_inference}---referred to as \textit{ordinary gradient descent} (OGD) within EFD--FEP models---is its modulation by the coefficient $G_{q_l}(\eta_{q_l})$. In information geometry, the Legendre-dual parameters $\eta$ and $\mu$ are treated as distinct coordinate systems within a shared space of distributions, known as a \textit{statistical manifold}. Generally, one coordinate system appears distorted from the perspective of the other, with the degree of distortion measured by the Fisher information matrix $G_{q_l}(\eta_{q_l})$.

The effect of this coefficient can hinder inference efficiency. Crucially, $G$ corresponds to the derivative of the relationship between $\mu$ and $\eta$, which we map to the activation function (or F--I curve) of each neuron in Section~\ref{sec:mapping}. For instance, consider a Bernoulli distribution where the activation function is sigmoidal. When the firing probability is sufficiently close to $0$ or $1$, the gradient of the activation function vanishes. As this is directly reflected in $G$, the update of the natural parameter $\eta$ becomes negligible when the neuron is confident in its current estimate.

To counteract this, information geometry proposes the \textit{natural gradient descent} (NGD), which applies the inverse Fisher information matrix to the dynamics \cite{Amari2016-dt}:
\begin{itembox}[l]{Inferential dynamics (natural gradient descent)}
\begin{equation}
    \tau_\text{inference} \dot{\eta}_{q_l} = -\eta_{q_l} + \eta_{p_l} + W_l^{\text{pred}~\top} \bar{\varepsilon}_l.
    \label{eq:ngd_inference}
\end{equation}
\end{itembox}
Under NGD, the dynamics simplify to this concise expression. This corresponds to updating $\eta_{q_l}$ in the steepest direction of the VFE as measured in the $\mu_{q_l}$-coordinates. Intuitively, by cancelling the distortion of the manifold, NGD ensures more rapid updates and, generally, faster convergence than OGD for functions defined on a statistical manifold.

Another advantage of NGD in the EFD--FEP model is its straightforward neural implementation. In general, $G$ contains non-zero off-diagonal components. If each dimension of $\eta_{q_l}$ is mapped onto a single neuron, OGD would require each neuron to integrate the states of all other neurons to update its own. In contrast, NGD updates can be performed via simple linear input reception, allowing the computation to be contained within a single neuron. This computational simplicity, combined with efficient inference, makes NGD an attractive candidate for biological implementation.

However, we do not necessarily claim that NGD is the sole biologically plausible mechanism. While $G$ may reduce convergence speed, it can also act as an \textit{evidence accumulator}. Biological neural systems are inherent with stochasticity and noise; instantly adapting to every received signal could lead to inferential instability. The choice between NGD and OGD can thus be viewed as an efficiency--stability tradeoff. Both regimes offer distinct advantages, particularly in distributions where $G$ is diagonal, making both plausible for biological systems depending on the functional requirements.

\subsubsection{Stochastic variants\label{sec:stochastic}}
This section introduces a stochastic variants of the inferential dynamics. The deterministic dynamics derived in Section~\ref{sec:derivation} is inherently based on the posterior mean $\mu_{q_l}$. However, biological neural systems are filled with randomness and stochasticity, making the message-passing of the exact deterministic mean values implausible. Instead of expressing inference in such a manner, here we introduce a stochastic dynamics based on non-deterministic firing activities.

Formally, this stochastic variant can be introduced by simply replacing the error term calculated as residuals between the actual lower-layer activities (or sensory inputs) and the uncertainty-weighted mean-based prediction with the prediction error based on the sampled realisation:
\begin{equation}
    \varepsilon_l:=\check x_{l-1}-G_{\phi_l}\xi_{\phi_l}=\check x_{l-1}-G_{\phi_l}W_l^\text{pred}\check x_l. \notag
\end{equation}
The resulting stochastic inferential dynamics introduced by this replacement is
\begin{itembox}[l]{Inferential dynamics (stochastic ordinary gradient descent)}
\begin{equation}
    \tau_\text{inference}\dot\eta_{q_l}=G_{q_l}(\eta_{q_l})\left\{-\eta_{q_l}+\eta_{p_l}+W_l^{\text{pred}\,\top}\varepsilon\right\},\label{eq:sogd_inference}
\end{equation}
\end{itembox}
and we refer to this expression as \textit{stochastic ordinary gradient descent} (SGD). We can apply the same replacement for NGD to obtain \textit{stochastic natural gradient descent}:
\begin{itembox}[l]{Inferential dynamics (stochastic natural gradient descent)}
\begin{equation}
    \tau_\text{inference}\dot\eta_{q_l}=-\eta_{q_l}+\eta_{p_l}+W_l^{\text{pred}\,\top}\varepsilon_l.\label{eq:sngd_inference}
\end{equation}
\end{itembox}

These stochastic variants are effective approximation of the deterministic inferential dynamics (Eqs.~\ref{eq:ogd_inference} and ~\ref{eq:ngd_inference}) in terms of expectation. Obviously, at each time step, the updates of the posterior parameter $\eta_{q_l}$ following Eqs.~\ref{eq:sogd_inference} and \ref{eq:sngd_inference} generally do not equal that following the corresponding deterministic expressions in Eqs.~\ref{eq:ogd_inference} and \ref{eq:ngd_inference}. However, when the timescale $\tau_\text{inference}$ of this dynamics is sufficiently faster than the timescale of changes in the sensory inputs $\check x_{0}$, the mean trajectories induced by SOGD and SNGD are considered to converge to their deterministic counterparts. These variants reflecting the stochastic nature of the biological neural systems can also reduce the VFE on average. 

\subsubsection{Learning Parameters with Local Plasticity Rules\label{sec:learning}}
Here, we demonstrate that parameter learning in this model can be expressed as local plasticity rules. By introducing an additional linearity assumption for the regulation of prior distributions—wherein priors are modulated by lateral (recurrent) inputs from the same layer $l$ and top-down inputs from the adjacent higher layer $l+1$—we identify three matrices of learnable parameters that facilitate the inferential dynamics: the prediction matrix and the two prior-regulating matrices. We derive the gradient of the VFE with respect to these three matrices and show that they all take simple forms involving only local signals in their update rules.

As a prerequisite, let $W_l^\text{rec} \in \mathbb{R}^{d_l \times d_l}$ represent lateral connections from neurons within the same layer, and $W_l^\text{TD} \in \mathbb{R}^{d_l \times d_{l+1}}$ represent top-down connections. We assume that these connections regulate the natural parameters $\eta_{p_l}$ of the prior distribution $p_l$ as:
\begin{equation}
    \eta_{p_l} = W_l^\text{rec} \mu_{q_l} + W_l^\text{TD} \mu_{q_{l+1}}, \label{eq:prior_regulator}
\end{equation}
which allows the dynamics (Eq.~\ref{eq:ogd_inference}) to be rewritten as:
\begin{equation*}
    \tau_\text{inference} \dot{\eta}_{q_l} = G_{q_l}(\eta_{q_l}) \left\{ -\eta_{q_l} + W_l^\text{rec} \mu_{q_l} + W_l^\text{TD} \mu_{q_{l+1}} + W_l^{\text{pred}\,\top} \bar{\varepsilon}_l \right\}.\label{eq:ogd_complete}
\end{equation*}
Note that the above corresponds to the deterministic OGD case. In SOGD, one may instead use $\check{x}_l$ and $\check{x}_{l+1}$ as the sources of prior regulation to obtain:
\begin{equation*}
    \tau_\text{inference} \dot{\eta}_{q_l} = G_{q_l}(\eta_{q_l}) \left\{ -\eta_{q_l} + W_l^\text{rec} \check{x}_l + W_l^\text{TD} \check{x}_{l+1} + W_l^{\text{pred}\,\top} \varepsilon_l \right\}.\label{eq:ngd_complete}
\end{equation*}
For NGD and SNGD, the coefficient $G_{q_l}(\eta_{q_l})$ is cancelled. Below, we discuss the gradient of the VFE with respect to these three matrices, focusing primarily on the OGD case. While the resulting gradients do not differ for NGD, appropriately replacing the variables ($\mu_{q_l} \rightarrow \check{x}_l$ and $\bar{\varepsilon}_l \rightarrow \varepsilon_l$) yields the learning rules for SOGD and SNGD.


First we derive the gradient of the VFE (Eq.~\ref{eq:EFD_VFE}) with respect to $W_l^\text{pred}$. Since this matrix appears only in the log-likelihood term, we can evaluate this gradient as
\begin{align}
    -\nabla_{W_l^\text{pred}}\mathcal F&=-\nabla_{W_l^\text{pred}}L \notag\\
    &=\check x_{l-1}\mu_{q_l}^\top-(G_{\phi_l}\bar\xi_{\phi_l})\mu_{q_l}^\top -G_{\phi_l}W_l^\text{pred}G_{q_l}(\eta_{q_l}) \notag\\
    &=\bar\varepsilon_l\mu_{q_l}^\top-G_{\phi_l}W_l^\text{pred}G_{q_l}(\eta_{q_l}).\label{eq:w_pred_grad}
\end{align}

Next, we derive the learning rules for $W_l^\text{rec}$ and $W_l^\text{TD}$. Since the mathematical structures of the learning rules for these two matrices are identical, we use $W_l^\bullet$ and $\mu_\bullet$ as placeholders. Unlike $W_l^\text{pred}$, these matrices appear only within the KL term; therefore, we evaluate the gradient of the KL term with respect to $W_l^\bullet$:
\begin{align}
    \nabla_{W_l^\bullet} D_\text{KL}[q_l \,\|\, p_l] &= \nabla_{W_l^\bullet} A_{p_l}(\eta_{p_l}) - \nabla_{W_l^\bullet} (\eta_{p_l}^\top \mu_{q_l}) \notag \\
    &= \mu_{p_l} \mu_\bullet^\top - \mu_{q_l} \mu_\bullet^\top \notag \\
    &= -(\mu_{q_l} - \mu_{p_l}) \mu_\bullet^\top. \label{eq:w_bullet_grad}
\end{align}

Combining Eqs.~\ref{eq:w_pred_grad} and \ref{eq:w_bullet_grad}, we obtain the following learning rules:
\begin{itembox}[l]{Learning rules}
\begin{align}
    \tau_\text{plasticity} \dot{W}_l^\text{pred} &= \bar{\varepsilon}_l \mu_{q_l}^\top -G_{\phi_l}W_l^\text{pred}G_{q_l}(\eta_{q_l}), \label{eq:learning_pred} \\
    \tau_\text{plasticity} \dot{W}_l^\text{rec} &= (\mu_{q_l} - \mu_{p_l}) \mu_{q_l}^\top, \label{eq:learning_rec} \\
    \tau_\text{plasticity} \dot{W}_l^\text{TD} &= (\mu_{q_l} - \mu_{p_l}) \mu_{q_{l+1}}^\top. \label{eq:learning_td}
\end{align}
\end{itembox}
As discussed in Section~\ref{sec:mapping}, these three learning rules correspond to local plasticity mechanisms. Under the EFD assumption, not only the inferential dynamics but also the learning rules can be expressed in these remarkably simple forms. Note that the update for $W_l^\text{pred}$ takes a slightly different form compared to the others, particularly concerning their neural implementation. However, we posit that each corresponds to distinct, biologically observed plasticity rules in different neurons or their respective compartments.

\subsection{Subtyping the model\label{sec:subtyping}}
In this section, before discussing the possible neural implementations of the EFD--FEP model, we define two subtypes based on their mathematical characteristics. While one subtype takes a more complex form, we argue that the other has a straightforward neural counterpart, potentially allowing for a mapping onto the laminar structures of the cerebral cortex.

First, we define \textit{Type-A EFD--FEP models} as those in which the log-partition function of the approximate posterior $q_l$ is additively decomposable:
\begin{align}
    A_{q_l}(\eta_{q_l}) = \sum_i a_{q_l, i}(\eta_{q_l, i}),
\end{align}
where $a_{q_l, i}: \mathbb{R} \rightarrow \mathbb{R}$ is a scalar-to-scalar function and $\eta_{q_l, i}$ is a scalar parameter. This condition is equivalent to the case in which $q_l$ can be factorised into independent univariate distributions:
\begin{align}
    q_l(x_l; \eta_{q_l}) = \prod_i q_{l, i}(x_{l, i}; \eta_{q_l, i}).
\end{align}

Type-A models possess several important characteristics. First, the Fisher information matrix $G_{q_l}(\eta_{q_l}) = \nabla^2_{\eta_{q_l}} A(\eta_{q_l}) = \partial \mu_{q_l} / \partial \eta_{q_l}$ is diagonal. Formally, this implies that a change in a single dimension of the $\eta_{q_l}$ space affects only the corresponding dimension in the $\mu_{q_l}$ space. Geometrically, while inhomogeneous values along the diagonal components may reflect distortions in point-to-point distances, the angles between any two basis vectors are guaranteed to be orthogonal in both coordinate systems. This orthogonality ensures that the angles between any two vectors in this space are preserved under parallel translation.

Second, the derivative of $a_{q_l, i}$ is strictly monotonically increasing. This is guaranteed by the requirement of strict convexity for $A_{q_l}$. Since convexity is a global characteristic of the EFD, the axis-wise convexity is preserved under the factorisability assumption.

We define \textit{Type-B EFD--FEP models} as those that do not satisfy the Type-A criteria. By definition, the characteristics discussed above do not apply to Type-B models. For instance, the coordinate systems are non-orthogonal, meaning that a translation along one axis affects multiple components of $\mu_{q_l}$, and angles between vectors are not preserved.

Intuitively, these subtypes reflect whether the dimensions of $\check{x}_l$ are independent within the approximate posterior. For example, independent Bernoulli or Poisson distributions—where only dimension-wise mean parameters determine the specific shape of the distribution—lead to Type-A models. This subtype also includes Gaussian distributions with diagonal covariance matrices. In contrast, general Gaussian distributions with non-zero off-diagonal covariance elements constitute Type-B models.

In Section~\ref{sec:mapping}, we discuss how Type-A models lend themselves to straightforward neural implementations. Although this does not generally hold for Type-B models—or more precisely, their implementation depends on specific interpretations—we also discuss how certain Type-B models might be supported by specific neural circuits or non-neuronal mechanisms.

\subsection{Biological mapping\label{sec:mapping}}
In this section, we discuss the potential biological implementations of EFD--FEP models. Specifically, we explore how neuronal properties may correspond to each variable appearing in the inferential dynamics and the learning rules.

The following discussion primarily focuses on the neural implementation of Type-A models. Due to the diagonality of their Fisher information matrix and the axis-wise independence of the gradient $\nabla_{\eta_{q_l}} A$, Type-A models lend themselves to a straightforward neural architecture. In Section~\ref{sec:typeb_implementation}, we contrast Type-B models with Type-A and exploratively discuss the conditions under which they might be implemented within neural systems.

\subsubsection{Representational and error-coding neurons}
First, we delineate the roles of representational and error-coding neurons in the implementation of Type-A models, following the conventions of standard PC theory. We assume a hierarchical neural network composed of distinct functional layers. We propose that the parameters $\eta_{q_l}$ and $\mu_{q_l}$ of the approximate posterior $q_l$ are implemented by the states of the \textit{representational neurons} in the $l$-th layer of the hierarchy. Conversely, the error signals $\varepsilon_l$ (or $\bar{\varepsilon}_l$) are represented by \textit{error-coding neurons}.

\subsubsection{Internal states and firing activities\label{sec:actfunc}}
In this section, we relate the mathematical variables appearing in the inferential dynamics and learning rules to physiological substrates. Each representational neuron is associated with two variables: $\eta_{q_l, i}$ and $\mu_{q_l, i}$. As demonstrated in Section~\ref{sec:efd}, these two parameters are dual via the log-partition function. We refer to $\eta_{q_l, i}$ as the \textit{internal state} of the neuron, while $\mu_{q_l, i}$ represents its \textit{firing rate}. Given that the firing rate is defined neurophysiologically as the mean firing activity, we picture that $\check{x}_{l, i}$ corresponds to the actual stochastic firing activity (e.g. individual spikes or instantaneous counts) of a single neuron.

While internal states may not have a single direct physical substrate, we argue they are closely related to the dynamical features of membrane potentials. A key characteristic of the relationship between $\eta_{q_l}$ and $\mu_{q_l}$—particularly in Type-A models—is that $\mu_{q_l}$ is a strictly monotonically increasing function of $\eta_{q_l}$. Thus, a straightforward interpretation is that the internal state regulates the firing rate in a positively correlated manner. We suggest that the biological counterpart of the internal state could be, for instance, the rate of membrane repolarisation following a spike-after-hyperpolarisation (undershoot). If the recovery speed increases, the expected spike frequency rises, and vice versa. While Eq.~\ref{eq:ogd_inference} shows that $\eta_{q_l}$ is regulated by dendritic inputs, we consider the internal state to be distinct from the raw input current; a sudden offset of input current does not necessarily cause an instantaneous cessation of firing, but rather a gradual decay in the firing rate.

Furthermore, the model contains an important attribute that determines firing patterns: the log-partition function $A$. In the EFD regime, $A$ dictates the specific shape of the distribution. We relate the gradient $\nabla_{\eta_{q_l}} A(\eta_{q_l})$ to the \textit{activation function} or the \textit{frequency--current (F--I) curve} of a neuron. Specifically, in Type-A models, the individual F--I curves correspond to each $a'_{q_l, i}(\eta_{q_l, i})$. This is consistent with traditional electrophysiological methodologies for observing F--I curves. Let $c$ be a constant input to a neuron, defined as $c = \eta_{p_l, i} + w_{l, i}^\text{pred}~\bar{\varepsilon}_l$, where $w_{l, i}^\text{pred}$ is the $i$-th column of $W_l^\text{pred}$. The attractor of the dynamics under this condition, $\tilde{\eta}_{q_l, i}$ such that $\dot{\eta}_{q_l, i} = 0$, satisfies $\tilde{\eta}_{q_l, i} = c$. Applying the activation function to both sides yields the steady-state relationship:
\begin{equation*}
    \tilde{\mu}_{q_l, i} = \nabla_{\eta_{q_l, i}} A_{q_l}(c),
\end{equation*}
which defines the mapping from constant input to firing rate. In electrophysiology, F--I curves are typically characterised by the steady-state firing rate achieved after a sustained period of constant-intensity input. The $c$--$\tilde{\mu}_{q_l}$ relationship derived above is thus a direct mathematical analogue of this experimental paradigm.

\subsubsection{Synaptic plasticity}
The learning mechanisms described in Eqs.~\ref{eq:learning_pred}--\ref{eq:learning_td} relate to local plasticity rules. Assuming these matrices correspond to synaptic weights, all three types of learning rules are regulated by the product of the pre-synaptic firing rates (or stochastic spikes in SOGD/SNGD) and specific post-synaptic states.

The plasticity of the prediction synapses (Eq.~\ref{eq:learning_pred}) can be interpreted as a standard Hebbian-like rule with uncertainty-weighted decay term. Since the pre- and post-synaptic elements of $W_l^\text{pred}$ are the representational neurons and the error-coding neurons, respectively, this formal expression reflects a fundamental principle: the strengthening of synapses between a simultaneously firing pre-post synaptic pair.

In contrast, the plasticity rules for the prior-regulating synapses $W_l^\text{rec}$ and $W_l^\text{TD}$ appear similar yet distinct. Unlike a simple correlation of neuronal activities, the post-synaptic term $(\mu_{q_l} - \mu_{p_l})$ represents the difference between the current firing rate and the rate dictated by the prior regulators. Notably, $\mu_{p_l}$ is the integrated sum of prior-regulating inputs (\textit{cf.} Eq.~\ref{eq:prior_regulator}). At first glance, this may seem biologically implausible, as an individual synapse typically lacks access to the total input from other synapses. However, there is a compelling potential correlate in neurophysiology. If we assume representational neurons are pyramidal cells, it is well established that inputs at the apical tuft are integrated before propagating to the soma. Furthermore, recent evidence suggests a mechanism where back-propagating action potentials or dendritic plateaus meet these integrated apical inputs \cite{Larkum1999-no, Larkum2013-tz}, triggering synaptic plasticity based on the voltage difference between the total input and the plateau \cite{Bittner2015-mp}. Based on this correspondence, we postulate that if cortical circuits implement the EFD--FEP model, representational neurons are indeed pyramidal cells, where lateral and top-down inputs target the apical dendrites, while bottom-up error-feedback signals are received at the basal dendrites.

\subsubsection{Heterogeneous networks}
An important feature of the EFD--FEP model—particularly in Type-A cases\footnote{Even when the model is not strictly Type-A, this holds if the approximate posterior can be decomposed into independent subfactors, each of which may be internally non-diagonal: e.g. where neurons 1--10 encode a non-diagonal Gaussian and neurons 11--20 encode a categorical distribution.}—is its capacity to embrace the electrophysiological heterogeneity of neurons. The fundamental property of the EFD that determines the specific shape of a distribution is the log-partition function. In Type-A models, the factorised components $a_{q_l, i}$ can be independently selected, provided they remain strictly convex. This allows each neuron to possess its own activation function $a'_{q_l, i}$ with characteristics distinct from those of other neurons within the same population. Even when a network consists of neurons with such heterogeneous response properties, they collectively construct a single approximate posterior, and the reduction of the VFE via the dynamics derived in Eq.~\ref{eq:ogd_inference} remains guaranteed.

This theoretical framework reflects the electrophysiological heterogeneity observed in biological neural networks. It has been demonstrated that even at meso- to microscopic scales, local networks are ensembles of neurons with diverse F--I curves; some exhibit sigmoidal responses while others show exponential profiles. Since the activation function in our model depends on the choice of the log-partition function, this diversity in response properties can be interpreted as a diversity of encoded posterior distributions. For instance, sigmoidal tuning may correspond to a Bernoulli or binomial distribution, whereas exponential tuning suggests the encoding of a Poisson distribution.

This represents a novel contribution to research on FEP. Previous theoretical and modelling studies have predominantly assumed Gaussian distributions, leading to the assumption of homogeneous response tunings—effectively identical projections. In contrast, our theory reveals that such homogeneity is not a necessary condition for VFE reduction via PC-like dynamics. Instead, heterogeneity may even enrich the representational capacity for estimating latent states.

\subsubsection{Type-B models\label{sec:typeb_implementation}}
While the straightforward neural mapping discussed above is primarily plausible under the Type-A assumption, we now consider the possibility that certain distributions within the Type-B regime may also be implemented in neural systems. 

A potential challenge for the brain in implementing a Type-B EFD--FEP model is that the computation of the coefficient $G_{q_l}(\eta_{q_l})$ can be extremely demanding. In Type-B models, $G$ depends on the current internal states of the entire neuronal population, typically in a complex, non-linear manner. This difficulty may be bypassed if neural populations encoding Type-B distributions utilise NGD or SNGD, where the $G$ term is effectively cancelled in the update rule.

Even under NGD, another issue arises regarding the nature of the activation function. Since the log-partition function cannot be decomposed in a component-wise manner in Type-B models, each component of the activation function $\nabla_{\eta_{q_l}} A(\eta_{q_l})$ necessarily involves the internal states or firing rates of other neurons. A prime example is the non-diagonal Gaussian distribution. Friston and his colleagues have long grappled with mapping these complex "activation functions" (though they do not explicitly use this terminology) onto neural systems.

However, for certain classes of distributions, seemingly implausible activation functions can be supported by known neurophysiological mechanisms. A notable example is the \textit{categorical distribution}. When a categorical distribution is assumed, the activation function for each neuron takes the form of the softmax function:
\begin{equation*}
    \mu_{q_l, i} = \frac{\exp(\eta_{q_l, i})}{\sum_j \exp(\eta_{q_l, j})}.
\end{equation*}
Although this form incorporates the states of all neurons in the population, its functional property is clear: a neuron's firing rate is suppressed when the internal states of other neurons are high, and increased when they are low. Possible neural correlates for this include \textit{lateral inhibition} and the resulting \textit{winner-take-all} circuitry. Alternatively, global changes in the potential of the local extracellular electrolyte could provide a non-synaptic mechanism for such collective inhibition. While we do not provide a universal mathematical condition for the implementability of Type-B models, we suggest that their biological counterparts should be explored on a case-by-case basis.

\section{Simulation with heterogeneous Bernoulli model\label{sec:hetero_bernoulli_res}}
This section shows that a concrete implementation of our EFD--FEP model can learn internal activities in response to the input sequences and appropriately reduce the VFE and prediction error along the sequence through an \textit{in sillico} demonstration. While the EFD is a broad class of probability distributions, we show specifically the implementation of this model under heterogeneous Bernoulli setting. In the heterogeneous Bernoulli setting, belonging to type-A models, a high-dimensional binary value $\check x_l=x_l\in\{0, 1\}^{d_l}$ sampled from the variational posterior $q_l$ at each time step is the activities of $d_l$ neurons in the layer $l$. Since these values are binary, the sampled values are seen as spikes. Hence, this model is a sampling-based spiking neural network.

\subsection{Time-discretisation of the model dynamics}
To simulate the dynamics of the model, we first time-discritised the SNGD inference dynamics (Eq.~\ref{eq:sngd_inference}). The particular way to discritise the dynamics is the Euler method. Let us now have the length $\Delta t$ of a discrete timestep. In the Euler method, the discrete-time dynamics at each time point $t$ is expressed as
\begin{equation}
    \eta_{q_l}(t+\Delta t)=\eta_{q_l}(t)+\frac{\Delta t}{\tau_\text{inference}}\big(-\eta_{q_l}(t)+\eta_{p_l}(t)+W_l^{\text{pred}\,\top}\varepsilon_l(t)\big).
\end{equation}
We denote $\tilde\tau_\text{inference}:=\frac{\Delta t}{\tau_\text{inference}}$ and set it to $0.1$.

Likewise, the same discretisation can be performed for the learning dynamics in Eqs.~\ref{eq:learning_pred}--\ref{eq:learning_td}. For the discrete version of the learning dynamics, we set the value $\tilde\tau_\text{plasticity}=0.002$.

\subsection{Heterogeneous Bernoulli setting}
In this demonstration, we set heterogeneous multivariate Bernoulli distribution as a concrete choice of variational posterior and prior. That is, $x_l\in\{0, 1\}^{d_l}$ and
\begin{equation}
    A_{q_l}(\eta_{q_l})=\sum_{i=1}^{d_l}\frac{1}{\alpha_{l, i}}\log\big(1+\exp[\alpha_{l, i}(\eta_{q_l, i}-\beta_{l, i})]\big),
\end{equation}
where $i$ indexes the dimension of the feature space. Since this log-partition is a sum of dimension-wise function, one can see that the EFD--FEP model under the heterogeneous Bernoulli setting is a type-A model. The derivative of each dimension-wise function is
\begin{equation}
    a'_{q_l, i}(\eta_{q_l, i})=\frac{1}{1+\exp[-\alpha_{l, i}(\eta_{q_l, i}-\beta_{l, i})]},
\end{equation}
which is a sigmoid function with the gain parameter $\alpha_{l, i}$ and the shift parameter $\beta_{l, i}$. The shift parameter determines the point where the firing probability is $1/2$: when it takes a large positive value, the neuron needs to reach high value of $\eta_{q_l}$ to achieve firing probability $1/2$. On the other hand, the gain parameter $\alpha_{l, i}$ determines how rapidly the neuron's firing changes in respond to changes in $\eta_{q_l}$. The rapidness rate is not uniformly determined, but rather depends on the current state $\eta_{q_l}$ When it takes a large positive value, the response is slow in most regime while it sharply changes its firing probability from near-zero to near-one value around the regime where $\eta_{q_l, i}\approx\beta_{l, i}$; conversely, for small gain values, the neurons tend to exhibit slow but not quasi-stationary changes of firing probability globally in the $\eta_{q_l, i}$ space, while the sharpness is weakened and the response rate is more uniform.

At initialisation of the model, we assigned different values of the gain and the shift to neurons at each layer of the network. The parameter values are sampled from uniform i.i.d.: $\mathbb R\ni\alpha_{l, i}\sim\text{Unif}(0.5, 3.5)$, $\mathbb R_+\ni\beta_{l, i}\sim\text{Unif}(1, 4)$. This introduces heterogeneous response properties in neurons in a population.

In this demonstration, we construct a network with two layers. The first layer has 512 neurons, while the second layer has 256 neurons. The second layer, since it does not have a higher layer, does not have the top-down prior-regulating synapses $W_2^\text{TD}$.

\subsection{Training}
Since our model can run the inferential dynamics (Eq.~\ref{eq:sngd_inference}) and the learning dynamics (discretised version of Eqs.~\ref{eq:learning_pred}--\ref{eq:learning_td}), it does not have separate training and inference phases. Given a sequence of inputs, the model can infer the latent state and predict the incoming inputs or the activities in the lower layer, while learning the appropriate connectivity. Particularly in our setting, we set the time constants $\tilde\tau_\text{plasticity} \ll\tilde\tau_\text{inference}$, which indicates that the inferential dynamics is sufficiently faster than the learning dynamics. We expect from this that the inferential dynamics capture the features of different images of the digits in a sequence, while the learning dynamics captures the overall data distribution in the whole dataset.

While the algorithmic implementation does not separate the learning and inference, we separate the training procedure from the testing procedure. In the training procedure, we feed the input sequences to the model and allow it to run both the inferential and the learning dynamics simultaneously. In the testing phase, we freeze the learned connectivities and analyse the internal neural activities given sequences sampled from a different subset of the dataset from the training procedure. 

The input sequences presented in the training procedure are sampled sequences from the MNIST dataset. In a sequence, the presented image changes every 20 timesteps and one sequence lasts for 240 timesteps. In each sequence, the order of the presented digits are incremental; if the sequence starts with the digit ``8'', then the following digits are ``9'', ``0'', ``1'', and the last digit would be ``9''. 

\subsection{Demonstrated neural dynamics}
\begin{figure}
    \centering
    \includegraphics[width=\linewidth]{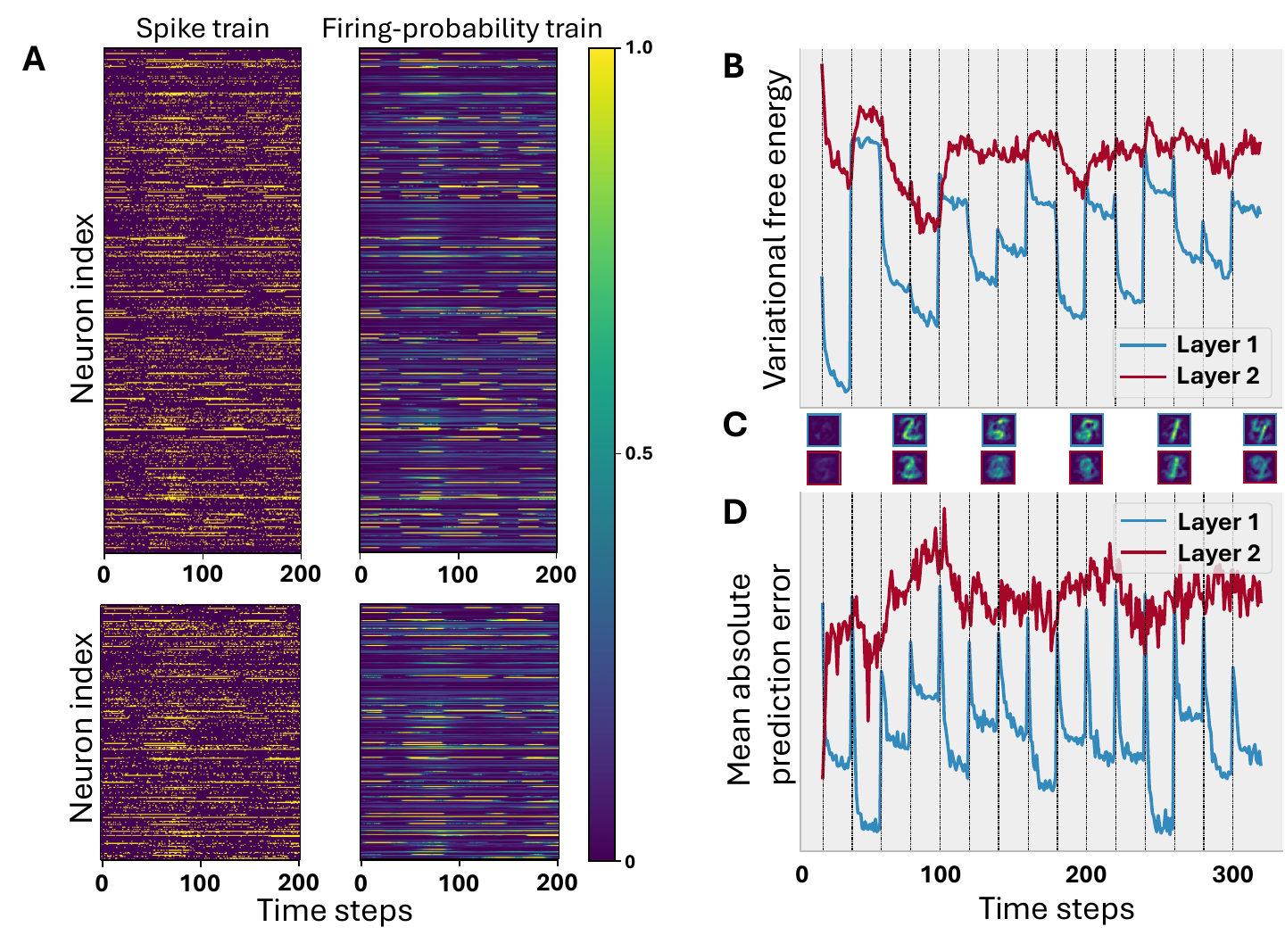}
    \caption{Simulation of EFD-FEP model under heterogeneous bernoulli setting. This simulation is done with sequential MNIST inputs, in which the presented digit changes every 20 timesteps. (A) The spike trains and firing probabilities (mean spikes) of neurons in the first and second layers in first 200 timesteps. The former corresponds to $\check x_l$ sampled from the variational posterior $q_l$ and the latter is visualisation of $\mu_{q_l}$, the expectation parameter of $q_l$. (B) VFE of each layer in the first 300 timesteps. The red and blue lines represent the VFE in the first and the second layers, respectively. The black dashed lines represent the timepoint where the input image switches. (C) The generated prediction of the input image from the first and the second layers at every 60 timesteps. (D) Absolute values of prediction errors averaged over the population at each timestep.}
    \label{fig:hetero_bernoulli_res}
\end{figure}
Fig.~\ref{fig:hetero_bernoulli_res} shows the activities of the demonstrated model in test procedure. These are the results from the testing procedure after 100 epochs of training procedure. In this particular sample of sequence, the first digit in the sequence is ``0''.

Fig.~\ref{fig:hetero_bernoulli_res}A shows an example of spike train (left panels) and firing-probability train (right panels) from the first (top panels) and second (bottom panels) layer. The horizontal and the vertical axes represent the timestep and neuron indices. For all of the four subpanels, one can qualitatively see that the firing pattern changes every 20 timesteps in response to the changes in the digit in the input sequence. There are some neurons that exhibit drastic changes of the firing rates and probabilities, while others tend to exhibit more ambiguous firing patterns that change relatively slowly compared to those rapidly responding neurons. Also, there is a slight tendency that the neurons in the second layer responds to the changes of the input more slowly compared to the second layer. The duration of a certain firing pattern tends to last longer than the first layer, while the first layer rapidly switches it to a different mode when the input pattern changes. Recall that the time constants for the inferential (and learning) dynamics do not differ between the first layer and the second. The layer-dependent difference of the response speed of the firing patterns despite this configuration may suggest that the network spontaneously acquired the coding of different timescales in the sequence, including the typical changes from a digit to another.

Fig.~\ref{fig:hetero_bernoulli_res}B shows that our model exhibits stable reduction of the VFE values. The horizontal axis of this panel represents the timesteps, while the vertical axis indicates the value of the VFE calculated at each layer and timestep. The blue and red lines are the VFEs in the first and the second layers, respectively. The black vertical dashed lines indicate the timesteps at which the input image changed. The blue lines, indicating the VFE at the first layer, tend to show increases followed by rapid decreases of the values at image-switching timepoints. After the rapid reduction for a several timesteps, the values tend to keep decreasing or stay at almost constant values. This indicates that the training procedure had achieved the connectivity with which the network can rapidly follow the changes of the inputs. The VFE at the second layer, shown in the red line, exhibited less variabilities during the sequence. Although the reduction rate of the VFE after changes in the input is lower than the first layer, so is the increase of the VFE at the image-switching timesteps; this implies that the neural representations of the inferred latent state in the second layer are more stable and invariant to the ordered changes in the input images.

It is noteworthy that the inferential dynamics we derived in the present article included an approximation that neglects the term related to the third cumulant of the variational posterior $q_l$. Instead of providing the concrete mathematical discussion on the validity of this approximation in Section~\ref{sec:derivation}, this demonstration, showing the constant reduction of the VFE, empirically confirms that it is a valid approximation of a dynamics minimising the VFE.

Fig.~\ref{fig:hetero_bernoulli_res}C shows the visualised prediction of the sensory inputs from the activities of the neurons in the first (top; in the blue box) and the second (bottom; in the red box) layers. The prediction from the first layer is calculated as $s_\text{pred}^{(1)}=G_{\phi_1}W_1^{\text{pred}}x_1$, while that from the second layer is $s_\text{pred}^{(2)}=G_{\phi_1}W_1^\text{pred}\left(G_{\phi_2}W_2^\text{pred}x_2\right)$. The prediction from the second layer then corresponds to ``the prediction from the first layer if the second layer accurately predicted the activity of the first layer''. First, one can qualitatively see that the prediction of the first layer is less ambiguous compared to the second layer. While both predictions are noisy due to the probabilistic sampling of the neuronal firings from the Bernoulli distributions, the first layer's prediction seems to capture clearer shape of the digits in the presented image. Although prediction from the second layer is more ambiguous, we argue that this does not necessarily mean that the second layer does not capture the information about the sensory input; rather, it may have maintain the inferential state within a region from which the system can respond to the changes of the inptus. In fact, at the 60th timestep (the second panel from the left), at which the input digit changes from ``2'' to ``3'', the prediction from the second layer looked rather close to ``3'' than ``2''. Recall that, since the input digit changes at this exact timepoint, the information that the input image changed had not necessarily reached the second layer. Yet, it exhibits a slight movement from the state predicting ``2'' to another predicting ``3''. Although such a behaviour was not observed at every change of the input images, a similar behaviour was also observed at the timestep 180 (third panel from the right), at which the input image has changed from ``8'' to ``9''. One can refer to Fig.~\ref{fig:hetero_bernoulli_res}B for these two characteristic timesteps to ensure that the rate of the VFE increases at them were less rapid than other image-switching timepoints.

Fig.~\ref{fig:hetero_bernoulli_res} shows the adaptation to the changing inputs from the perspective of the absolute prediction error averaged over all neurons at each layer: $d_l^{-1}\sum_i|\varepsilon_{l, i}|$. Similarly to the VFE behaviour shown in Fig.~\ref{fig:hetero_bernoulli_res}B, the prediction errors exhibited rapid increases followed by the rapid decreases at the image-switching timepoints. However, the decreases were typically faster than the decrease of the VFE. This is a confirmation that the VFE is not the prediction error itself, and rather it is a combination of different metrics including the prediction error and the entropy of the current state of the variational posterior.

\section{Discussion}
In this work, we have formally demonstrated that the correspondence between the free-energy principle (FEP) and Predictive Coding (PC) holds generally when the posterior, prior, and prediction distributions belong to the Exponential Family of Distributions (EFD). Based on this EFD assumption, we proposed a novel model of neural computation that implements Variational Free Energy (VFE) descending dynamics. The resulting inferential dynamics take a remarkably simple form, where bottom-up inputs (prediction errors) and lateral/top-down inputs (prior regulations) are integrated linearly to update the internal state; the Legendre transformation of this state then yields the observed firing rates. Furthermore, we showed that the learning rules are expressed as biologically plausible plasticity mechanisms. We argued that their physiological implementations may be realised within the basal and apical dendrites of pyramidal neurons.

Crucially, neural networks based on our theory surpass previous studies exploring the biological implementation of FEP by accommodating the electrophysiological heterogeneity of F--I properties. Such heterogeneity is a hallmark of biological neural systems. From a computational perspective, we suggest that this diversity is indispensable for enriching the representational capacity of complex probability distributions, thereby facilitating robust adaptation to sensory inputs from the external world.

\subsection{Geometric Interpretation of Inferential Dynamics}
The inferential dynamics of our model exhibit a remarkably parsimonious form. All inputs to a single neuron are linearly integrated within the $\eta_{q_l}$-coordinate system, and the Legendre transformation—implemented via the activation function—determines the mean firing activity.

This elegant expression provides geometric intuition regarding how distinct types of information are represented within the state space. A significant characteristic of these dynamics is that they constrain each input vector to a specific linear submanifold. For instance, the driving vectors originating from prediction-error feedback are confined to the subspace defined by $\text{span}(W_l^{\text{pred}\top})$ and possess no components orthogonal to it. If the synaptic weights are sufficiently sparse, this constraining submanifold does not span the entire statistical manifold of neural states. Consequently, if the submanifolds $\text{span}(W_l^\text{rec})$ and $\text{span}(W_l^\text{TD})$, which constrain the prior-regulating drives, possess components orthogonal to the prediction-error subspace, these components can be interpreted as conveying information that is inherently unexplainable by bottom-up signalling alone.

This geometric separation may correspond to our internal ``interpretation'' of the external world---for example, the inference of 3D spatial arrangements from 2D retinal projections. Investigating the geometric structure of how different informational streams are represented may yield a deeper understanding of how the brain integrates sensor-based bottom-up data with high-level prior beliefs.

\subsection{Potential Connection to Information Thermodynamics}
Our model possesses a potential connection to information thermodynamics via the mathematical equivalence between the EFD and the canonical distribution. In statistical mechanics and thermodynamics, the canonical distribution is employed to describe the distribution of states in a thermal system at equilibrium:
\begin{equation*}
    f(x) = \frac{1}{Z(\beta)} \exp(-\beta E(x)),
\end{equation*}
where $\beta = 1/k_B T$ is the inverse temperature and $E(x)$ represents the internal energy. Rewriting the log-partition function as $A(\beta) = \log Z(\beta)$, we obtain an expression mathematically identical to the EFD, where the internal state $\eta$ corresponds to the negative inverse temperature $-\beta$, and the sufficient statistics $T(x)$ correspond to the energy $E(x)$. 

Under this analogy, we can identify thermodynamic counterparts within our model. In fact, there has been a theoretical attempts to connect information processing with thermodynamics given this analogy \cite{Balsubramani2025-kz}. Recent advances in information thermodynamics have revealed that the minimisation of energy dissipation corresponds to the maximisation of predictive power \cite{Still2012-jv, Ueltzhoffer2020-vb}. This thermodynamic analogy may facilitate future discussions regarding the energetic efficiency of neural systems and the potential relationships between metabolic costs and the accuracy of inference.

It is worth noting, when establishing this direction, although EFD is mathematically equivalent to the canonical distribution, it might not be appropriate to treat the EFD--FEP model as an equilibrium system. While the canonical distribution characterises the equilibrium states, in the case of our model, the natural parameter (inverse temperature) keeps fluctuating under the effect of stochastic realisations of EFD posteriors at other layers, through the bottom-up prediction-error feedbacks $\bar\epsilon_l$ and top-down signals regulating the prior $p_l$. This corresponds to a thermal system that is instantaneously in local equilibrium, while exhibiting effective non-equilibrium behaviour due to fluctuations in the temperature parameter. Such systems have been discussed as superstatistical systems \cite{Beck_2004}, in which the effective statistics is well established for the case where the inverse temperature follows a stationary distribution. Yet, this framework is not directly applicable to our model; future works should further formalise a more complex case where the fluctuation of the natural parameter is given by a non-stationary source.

\subsection{The Privileged Status of Error-coding Neurons}
In this section, we critically examine the requirement for error-coding neurons within our model to possess specific, biologically demanding properties to ensure neural implementability. These error-coding neurons, which emerge from the mathematical necessity of evaluating the log-likelihood gradient, exhibit characteristics distinct from representational neurons:
\begin{enumerate}
    \item They possess neither intrinsic internal states nor independent temporal dynamics;
    \item They are assumed to propagate bottom-up prediction error signals via analogue (non-spiking) signals that can take negative values;
    \item Synaptic connectivity from lower-layer representational neurons to error-coding neurons is assumed to be a one-to-one, "hard-wired" structure;
    \item Synapses from error-coding neurons to higher-layer representational neurons are assumed to be perfectly symmetric to their counterparts.
\end{enumerate}
These attributes are often incompatible with our standard understanding of biological neural networks. Such distinctions are typically introduced in an \textit{ad-hoc} manner, which may undermine the biological plausibility of the implementation.

The issue of negative values (Property 2) can be addressed by architectural modifications. A common solution is to split error-coding neurons into distinct excitatory and inhibitory populations. By assuming rectification in the input--output relationship of each neuron, we can constrain the output synapses to either a strictly positive (excitatory) or negative (inhibitory) regime.

However, such a modification introduces further complications; an input from a single pre-synaptic representational neuron would need to reach the dendrites of a specific excitatory--inhibitory pair of error-coding neurons with equal weights but inverted signs. This "hard-wired" requirement is not only biologically implausible but also risks violating Dale’s Law, which states that a single neuron typically exerts the same chemical effect at all of its synaptic connections (i.e. either excitatory or inhibitory).

The symmetry issue (Property 4) has a potential solution in the form of \textit{feedback alignment}. Originally proposed in the machine learning literature, this technique suggests that asymmetric synapses between two neuronal populations can converge toward a symmetric structure if their update rules include decay terms and share the same statistical structure. Specifically, in our model, the update rule for top-down synapses from representational to error-coding neurons is $\dot{W}_l^\text{pred} \propto \bar{\varepsilon}_l \mu_{q_l}^\top$. Feedback alignment proposes that if the prediction connectivity includes a decay term:
\begin{equation*}
    \tau_W \dot{W}_l^\text{pred} = \bar{\varepsilon}_l \mu_{q_l}^\top - \lambda W_l^\text{pred}
\end{equation*}
and the feedback connectivity follows a symmetric update rule:
\begin{equation*}
    \tau_W \dot{W}_l^\text{FB} = \mu_{q_l} \bar{\varepsilon}_l^\top - \lambda W_l^\text{FB},
\end{equation*}
then the two weight matrices empirically converge to become the transposes of one another. While this remains an empirical observation, it provides a plausible mechanism for the emergence of symmetry in our model.

The remaining challenges lack straightforward solutions at present. A more refined mathematical formulation of the EFD--FEP model may be required to eliminate these unnatural assumptions. For instance, a re-factorisation of the layer-wise VFE (Eq.~\ref{eq:factorised}) that explicitly incorporates "error-estimating distributions" might yield a different form of inferential and learning dynamics. Such attempts to achieve more sophisticated neural implementability remain a vital direction for future research.

\section*{Acknowledgement}
This study is supported by JSPS KAKENHI Grant Number 26KJ0370. We also acknowledge the computational resources provided by the Scientific Computing and Data Analysis Section of Core Facilities at Okinawa Institute of Science and Technology.

\bibliography{biblio.bib}

\appendix
\section{On third cumulant-neglecting approximation}
In Section~\ref{sec:derivation}, we introduced an approximation where we neglect the third cumulant $\nabla_{\eta_{q_l}}^3A_{q_l}(\eta_{q_l})$ of the posterior distribution to establish the neural implementation of the inferential dynamics in terms of predictive coding. The approximation error is:
\begin{equation}
    \Delta(\eta_q):=\frac{1}{2}\nabla_{\eta_q}\text{Tr}\left[G_{\phi_l} W_l^\text{pred}G_{q_l}(\eta_{q_l})W_l^{\text{pred}\,\top}\right].
\end{equation}
Here, for simplicity, we define $M:=W_l^{\text{pred}\,\top}G_{q_l}(\eta_{q_l})W_l^{\text{pred}}$, with which we rewrite this approximation error as
\begin{equation}
    \Delta(\eta_{q_l})=\frac{1}{2}\nabla_{\eta_{q_l}}\text{Tr}[MG_{q_l}(\eta_{q_l})].
\end{equation}

With Einstein summation convention, the concrete expression of $k$-th dimension of the gradient is
\begin{align*}
    \Delta_k(\eta_{q_l})&=\frac{1}{2}\partial_k\Big(M^{ij}G_{q_l}(\eta_{q_l})_{ij}\Big)\notag\\
    &=\frac{1}{2}M^{ij}\Big(\partial_kG_{q_l}(\eta_{q_l})_{ij}\Big).
\end{align*}
Denoting $T_{ijk}:=\partial_kG_{q_l}(\eta_{q_l})_{ij}=\partial_k\partial_i\partial_jA_{q_l}(\eta_{q_l})$, we obtain a simple form of the approximation error as
\begin{equation}
    \Delta_k(\eta_{q_l})=\frac{1}{2}M^{ij}T_{ijk}.
\end{equation}

From a geometrical interpretation, this form of the approximation error suggests that it is determined by the coupling between the likelihood variance and the posterior skewness. $M=W_l^{\text{pred}\,\top}G_{\phi_l}W_l^\text{pred}$ measures the covariance matrix $G_{\phi_l}$ of the likelihood function pulled back in the posterior space. When considering a fixed-covariance Gaussian likelihood and assuming that the learning timescale $\tau_\text{plasticity}$ is sufficiently slower than the inference dynamics $\tau_\text{inference}$, this can be seen constant. The third-order tensor $T_{ijk}=\partial_i\partial_j\partial_kA_{q_l}$, on the other hand, is the third-order cumulant of the posterior distribution. In the main content, we assumed that the absolute values of this tensor is sufficiently small and neglected it to obtain the PC form as an approximation of the true gradient flow. Intuitively, this tensor measures the skewness of the posterior, or the divergence of its log-partition $A_{q_l}$ from the quadratic form plus affine terms, \textit{i.e.} non-Gaussianity of the posterior distribution. 

This does not necessarily guarantee that the VFE keeps reduced under this approximation in general. Our extension of FEP to a broader class, the EFD, essentially violates the non-Gaussianity of the posterior distribution, and we are even motivated to do so. This means $T_{ijk}$ can take large values depending on the choice of the posterior shape $A_{q_l}$ and result in inadmissible approximation error. For example, while the maximum third cumulant of the Bernoulli distribution is generally smaller than the second, any $n$-th cumulant of the Poisson distribution occupies equal order $e^{\eta_{q_l}}$ and exponentially grows for large $\eta_{q_l}$ values.

However, $M^{ij}$'s may counteract the non-Gaussianity and avoid rapid growth of the approximation error. As seen in Eq.~\ref{eq:learning_pred}, the plasticity rule for the prediction synapses contain the uncertainty-weighted decay term. From this, the absolute values of elements of $W_l^\text{pred}$ have strong tendency to remain small, especially when the learning dynamics is converged and prediction error is small, or the pre-synaptic firing activity $\mu_{q_l}$ (or $\check x_l$ in the stochastic setting) is weak. This effect is considered to work as a tendency to attract the approximation error to small-valued regime.

\section{Beyond Gaussian likelihood\label{apsec:nongaussian_likelihood}}
This section attempts to discuss the effect of a further relaxation where the likelihood function, as well as the posterior and the prior, is assumed to be non-Gaussian EFD. In the main body of the present article, we assumed that the posterior and the prior are in the general form of the EFD, while the likelihood function is left Gaussian. This partial relaxation from the typical settings having been taken in the previous studies derived the extended correspondence between the VFE minimisation and predictive coding. We argue that a further generalisation where the likelihood also takes a off-Gaussian form does not necessarily correspond to predictive coding, and discuss the extent to which predictive coding can serve as an approximation of the VFE descent under such generalisation.

Note that this section takes a different notation from the main content. First, we omit layer indices in the following discussion. Since this ambiguates the difference between the internal representations $x_{l}$ and the observation or the lower-layer activity $x_{l-1}$, here we denote the former as $z$ and the latter as $x$. We also assume that the sufficient statistics of the random variables are equal to the random variables themselves: $T(x)=\check x=x$ and $T(z)=\check z=z$. These notations are taken just for the sake of the simplicity of the following discussions, and one can apply the same discussions for the broader case, for example, in which the sufficient statistics are not the same as the random variables.

Let us now consider the case where the likelihood function belongs to the EFD, but is not necessarily Gaussian. Under this general assumption, the log-partition function $A_\phi$ of the likelihood function does not take a quadratic form as in Eq.~\ref{eq:quadratic_logpart}. Hence, its gradient takes a different form from Eq.~\ref{eq:ll_grad}:
\begin{align}
    -\nabla_{\eta_q}\mathbb E_q[A_\phi(\xi_\phi)]&=-\int\nabla_{\eta_q}q(z; \eta_q)A_\phi(\xi_\phi)dz\notag \\
    &= -\int q(z;\eta_q)\left(\nabla_{\eta_q}\log q(z; \eta_q)\right)A_\phi(\xi_\phi)dz\notag \\
    &=-\int q(z;\eta_q)(z-\mu_q)A_\phi(\xi_\phi)dz \notag \\
    & =-\mathbb E_q[(z-\mu_q)A_\phi(\xi_\phi)].\label{apeq:scorefunc_identity}
\end{align}
By plugging Eq.~\ref{apeq:scorefunc_identity} to the total log-likelihood gradient, it the prediction-error feedback looks different:
\begin{equation}
    -\nabla_{\eta_q}L=G_q(\eta_q)W^\top x-\mathbb E_q[(z-\mu_q)A_\phi(\xi_\phi)]\label{apeq:ll_grad}
\end{equation}

Here, we argue that this form of log-likelihood gradient feedback (Eq.~\ref{apeq:ll_grad}) does not have a straightforward counterpart neural implementation. This is because the log-partition gradient in Eq.~\ref{apeq:scorefunc_identity} has to be implemented as a unbiased estimator of mean of a complex nonlinear function. This unbiased estimator does not have an analytically closed form except for an empirical averaging; similarly to the SOGD and SNGD implementations discussed in Section~\ref{sec:stochastic}, replacing the whole feedback with
\begin{equation*}
    -\nabla_{\eta_q}L\approx G_q(\eta_q)W^\top x-(z-\mu_q)A_\phi(\xi_\phi)
\end{equation*}
and take a sufficiently faster time discretisation than the time scale of the sensory inputs, the averaged trajectory may approximate the true gradient flow. 

Even if the empirical average can serve as a close approximation of the true mean, another concern is that the term $(z-\mu_q)A_\phi(\xi_\phi)$ is considered biologically implausible. Intuitively, this form can be seen as a correction of the state of the representational neurons based on the deviation of the generated spikes from their current expectations with weight $A_\phi(\xi_\phi)$ regulated by the likelihood function; if a generated spike activity is stronger than expected, the state $\eta_q$ is suppressed, and \textit{vice versa}. However, the weight $A_\phi(\xi_\phi)$ is a function of $\xi_\phi=Wz$, which means that correction for the single-neuron state requires each neuron to know the current spike activity of all other neurons in the same layer as well as the synaptic weights from them. These properties, the instantaneous lateral communication between neurons and information sharing of the global synaptic weights are both biologically implausible.

Given this discussion, under the non-Gaussian likelihood assumption, biological neural systems can perform the VFE minimisation only in terms of approximation. Obviously, there can be various approximation strategies and predictive coding is then not the only option, but there can be other implementations whose performance could even exceed the PC-based approximation. Current trends around the FEP, including the present article, tend to discuss the VFE minimisation for sensory processing and perceptual inference under correspondence with predictive coding, but this discussion reveals that it is not necessarily an only option and surrogate approximations could lead to other mechanisms which have been observed in biological systems. Here, it is left an open question what kind of neural implementation can serve as valid approximation. This question can connect FEP to further biological context beyond predictive coding.

In the following, we argue the strategy to validate any neural implementation as approximation. The true gradient of the VFE is, according to the discussion above, in the following form:
\begin{equation}
    -\nabla_{\eta_q}\mathcal F=G_q\left\{-\eta_q+\eta_p+W^\top x\right\}-\mathbb E_q[(z-\mu_q)A_\phi(\xi_\phi)].\label{apeq:full_true_gradient}
\end{equation}
Now we take a surrogate dynamics that is neural-implementable,
\begin{equation}
    \dot\eta_{q}\propto G_q(\eta_q)\left\{-\eta_q+\eta_p+W^\top x\right\}+v,\label{apeq:surrogate}
\end{equation}
where $v$ is a surrogate gradient of the mean log-partition term. Even if it is not possible to construct $v$ that exactly equals Eq.~\ref{apeq:full_true_gradient}, the surrogate dynamics in Eq.~\ref{apeq:surrogate} can be seen as a valid approximation if
\begin{equation}
    -\nabla_{\eta_q}\mathcal F^\top\dot\eta_q\geq0\label{apeq:validity_checker}
\end{equation}
is satisfied, or,
\begin{equation}
    -\mathbb E_q[\nabla_{\eta_q}\tilde {\mathcal F}^\top\dot\eta_q]\geq0
\end{equation}
is, in case $v$ is a random variable dependent on the sampled spiking activity $z$ and we consider a stochastic VFE obtained by replacement $\mathbb E_q[(z-\mu_q)A_\phi(\xi_\phi)]\rightarrow(z-\mu_q)A_\phi(\xi_\phi)$. 

The validity checker in Eq.~\ref{apeq:validity_checker} can be evaluated as
\begin{align}
    (v-\mathbb E_q[(z-\mu_q)A_\phi(\xi_\phi)])^\top(-\nabla_{\eta_q}\{D_\text{KL}+\bar\xi_\phi^\top x\})+v^\top(-\mathbb E[(z-\mu_q)A_\phi(\xi_\phi)])\notag\\
    \geq-\|\nabla_{\eta_q}D_\text{KL}\|^2-\|\nabla_{\eta_q}\bar\xi_\phi^\top x\|^2+2(\nabla_{\eta_q}D_\text{KL})^\top(\nabla_{\eta_q}\bar\xi_\phi^\top x)\label{apeq:detail_validity_checker}.
\end{align}
If this inequality is satisfied in general, the surrogate dynamics is ensured to slowly reduce the VFE along its trajectory even if it is not the steepest gradient descent.

\end{document}